\documentclass[aps,pra,twocolumn,groupedaddress,showpacs]{revtex4}
\usepackage{epsfig}
\begin{document}

\title {The path and the multi-teaching issues in the coupled pendulum and
  mass-spring experiments } \author {I. Boscolo,  L. Gariboldi and
  R. Loewenstein \\ { \em University and INFN, via Celoria 16, 20133 Milano,
    Italy } }

\begin{abstract}
The main task of an introductory laboratory course is to foster students' manual, conceptual and statistical ability to investigate physical phenomena. Needing very simple apparatus, pendulum experiments are an ideal starting point in our first-year laboratory course because they are rich in both physical content and data processing. These experiments allow many variations, e.g. pendulum bobs can have different shapes, threads can be tied to a hook at their edge or pass through their centre of mass, they can be hanged as simple or bifilar pendulums. In these many variations, they emphasize the difference between theory and practice in the passage from an idealized scheme to a real experimental asset, which becomes evident, for example, when the pendulum bob cannot be considered an idealized point mass. Moreover, they require careful observation of details such as the type of thread used and its rigidity or the bob initial slant, which leads to different behaviors. Their mathematical models require a wide range of fundamental topics in experimental data analysis: arithmetic and weighted mean, standard deviation, central limit theorem application, data distribution, and the significant difference between theory and practice. Setting the mass-spring experiment immediately after the pendulum highlights the question of resonance, revises the gap between theory and practice in another context, and provides another occasion to practice further techniques in data analysis.
 \end{abstract}

 \maketitle

 \section {Introduction }

Many fundamental issues of a first year laboratory course are at stake when we use the simple pendulum \cite{book} experiment to measure {\it g}. Undergraduate texts explains the {\it theory} of an idealized scheme with a point mass attached to an ideal thread oscillating with very small amplitudes to justify the mathematical approximation used in the deduction: see Fig. \ref{pendoli1} scheme (a). In {\it practice}, a point mass does not exist, neither does an ideal thread, nor do infinitely small oscillations. Hence the problem of the passage from an idealized physical system to a real one \cite{2} must be discussed with students and the limits of a mathematical approximation must be assessed. Therefore in the laboratory we start the trial experiment by carefully engineering a systematic error in the first rough measures of {\it g} to emphasize the important difference between theory, which considers a point mass, and the laboratory experiment, which uses an extended body. After that we compare the parameters of two distinct normal distributions by asking students to measure fifty times, say, five and subsequently ten complete oscillations. The data form two gaussian distributions with compatible means but different dispersions. The analysis of these data require: (i) identification of a gaussian distribution, (ii) comparison of gaussian mean using the central limit theorem, and (iii) comparison of standard deviations. In the final part of the experiment students observe damping oscillations and check their mathematical model. Here they have to use their judgment to measure the time constant with two different methods and check the compatibility of the results obtained.

In the wider outline of a first laboratory course based on oscillations and waves, this experiment is followed by a second set of two four-hour laboratory sessions with a mass-spring experiment. In this new context students practice: (a) the passage from idealized theory to real laboratory conditions, in this case from a theoretical zero-mass spring to a spring with a relevant mass \cite{5}; (b) a straight-line fit; (c) observing damping and determining the more demanding resonance curve and its FWHM.


\section{General teaching procedure}

Introductory lectures present an explanation of theory and an outline of the experiment. Through discussions we define the variables to be measured, present suitable techniques and procedures, indicate some common errors, and give general advice. We ask students to write a rough plan of the experiment in advance.
In the laboratory students assemble the apparatus and follow their own plan. Assistance is given in terms of discussions and help to slower groups so that most of them proceed together through the various phases of the experiment. We interrupt the work at critical points to ask students to record on the board a summary of data from their lab notebooks for general statistical treatment and plenary discussion.


  \section{From theory to practice}

In the introductory lecture we start by discussing how to measure {\it g} in our laboratory. We first ask which is the most convenient physical law for that purpose, from the point of view of ease and precision. We are at the beginning of a physics course, so we compare the advantages and disadvantages of the law of free fall and the pendulum \cite{3}. In the end of the discussion it is agreed that the law describing the period of a simple pendulum \cite{4}

\begin{equation}
\label{eq-periodopendolo}
T= 2 \pi \sqrt{\frac{\ell}{g}}
\end{equation}
is the most convenient.
\\

We guide the next decision to prepare the ground for the comparison between two different normal distributions. We ask students to make two sets of measurements of the period. They measure fifty times the time required for five complete oscillations and then repeat the procedure for ten periods. Here they practice comparing distributions with compatible means but different dispersions.

Since the thread length $\ell$ and the period {\it T} can be measured with a precision of a few parts per thousand, it will be possible to detect the effects of very small systematic errors caused by thread rigidity \cite{7}\\ and the choice of initial conditions.


\section {Initial set-up and first measurements}

The first practical decision regards the length of thread to be used. After considering many possible values,  we decide to choose the same length for all groups for statistical reasons, say, 500 {\it mm}, on the grounds of not only minimizing relative errors but also having periods of oscillation that are not too long. Students are then asked to make a trial experiment to test the apparatus by measuring period {\it T} three times using each time five complete oscillations. The corresponding calculated value for {\it g} has to fall within a few percentage points of its expected value. In this initial phase they check that the apparatus is working properly, practice observing and timing techniques, and on completion copy a summary of data and results on the board, such as shown in the following table, for a brief plenary discussion.
\vspace{0.3cm}

\begin{tabular}{|c|c|c|c|c|c|c|c|c|}
\hline
no. & $\ell$ (cm) & T (s) & $\Delta g (trial) \%$ &
$\Delta g (trial)_{\ell+\Delta \ell} \%$  \\ \hline
1  & 50.4 & 1.463 & - 2 \%   &  9.76; $\Delta= - 0.5 \%$  \\ \hline  
2  & 50.3 & 1.453 & - 1.7 \% & 9.89; $\Delta= + 1 \%$   \\ \hline
3  & 49.9 & 1.452 & - 1.7 \% & 9.77; $\Delta= - 0.4 \%$  \\ \hline    
4  & 50.0 & 1.455 & - 2.1 \%   & 9.79; $\Delta= -0.1 \%$  \\ \hline    
5  & 50.2 & 1.454 & - 1.7 \% & 9.83; $\Delta= + 02 \%$  \\ \hline
6  & 50.0 & 1.448 & - 1.5 \%   & 9.86; $\Delta= + 0.5 \%$  \\ \hline    
7  & 50.2 & 1.450 & - 0.5 \%   & 9.84; $\Delta= - 0.04 \%$  \\ \hline    
\end{tabular}
\vspace{0.3cm}

Students notice that in the 4th column the values of {\it g} are all smaller than the expected value. We remark that measurement theory predicts that casual errors cause values to fall randomly on either side of the expected value, so through discussion we are forced to conclude that there is a systematic error. At this stage we consider relative errors and what the theoretical variables {\it T} and $\ell$ really represent in practice. The measurement of the period {\it T} is quite precise. We estimate its error and see that the problem must be the length $\ell$, which results shorter than its required value for the accepted {\it g} value. The first conceptual problem to face about the length is that the pendulum equation refers to a point mass while in our real set-up the mass is extended.

We have used round disks many times, but here we will discuss the use of conic bobs. By using a cone we avoid the fact that some students with previous lab experience anticipate the discussion considering that the pendulum length is the distance from the pole to the disk center. With the cone a similar consideration does not readily occur, and students, especially those coming from a classical background or with no previous lab experience, accept more easily the theoretical statement that the thread length corresponds to the variable $\ell$.\\

The class is guided through the following path: (1) the motion equation in books refers to a point mass, therefore the pendulum length $\ell$ is the thread length; (2) we realize that we have a systematic error and therefore a significantly different physical system; (3) we will make an alternative physical hypothesis; and (4) we will check the new hypothesis experimentally.
From the expected value of {\it g} it comes out $\Delta \ell \sim 20 \; mm$. This additional length corresponds to a point at 3h/4 where h is the height of the cone. We note that this distance from the top of the cone refers to the position of the center of mass. \\ 

So, given that it is an extended object, our system cannot move according to the point mass equation, $\vec{F}= m\, \vec{a}$. Its motion is described by the cardinal equations of a rigid body; in particular, we are dealing with the motion of the center of mass of the object. Repeating the calculation, assuming the pendulum length to be the distance from the pole to the center of mass of the object, we get the data in column 5, which are randomly distributed around the expected value \cite{6}. 
Considering students' average lack of lab experience, we guide the next decision to prepare the ground for the comparison between two different normal distributions. Having decided to consider $\ell$ to the center of the mass of the pendulum, students then proceed with the fifty measurements of 5{\it T} and 10{\it T} and obtain the two corresponding gaussian distributions. Here they practice comparing two normal distributions with compatible means but different dispersions. 

\begin{figure}[!ht]
\centering
\mbox{\epsfig{file=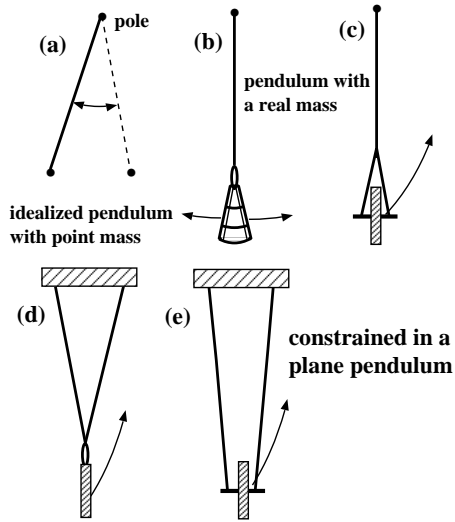,width=6cm}}
\caption{\label{pendoli1} Theoretical simple pendulum (a) and practical versions (b), (c), (d), and (e). }
\end{figure}

In the past few years we have tried the following variations illustrated in Fig. \ref{pendoli1}.


\section{Data and discussion}

From all the data regarding both distributions recorded in their notebooks the students copy on the board only the two means and the two standard deviations of {\it T} and their calculated value of {\it g}. One of the most important parts of this session is the plenary discussion of the parameters of these two distributions. Ideally the mean values for 5{\it T} and 10{\it T} ought to be compatible within each laboratory group because the thread length is exactly the same in both cases. In addition, the standard deviation of $T_{10}$ should be smaller than the standard deviation of $T_5$. Students are asked to check whether this happens with the results reported by each group on the board. Any possible discrepancies are discussed by the demonstrators. Given that the measured thread lengths cannot be exactly the same in different laboratory groups, we discuss with students whether all average values of $T_5$ or $T_{10}$ or {\it g} ought to be compatible. An encouragingly large number of students arrive at {\it g} as the only possible compatibility after calculating its result using the weighted means for the values of the period. A typical value for the relative error of $\bar{T}$ is around a fraction of percent and the corresponding error of $\ell$ is quoted at 1 or 2 {\it mm}, that is 2 or 4 per thousand. Hence, the relative error of {\it g} has to be in the range of the per thousand.

\begin{figure}[!ht]
\centering
\mbox{\epsfig{file=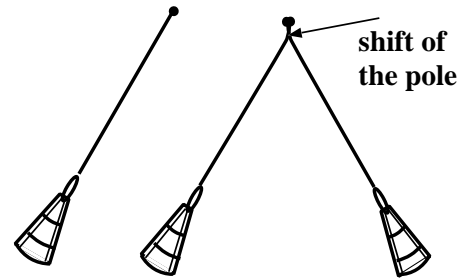,width=6cm}}
\caption{\label{piega-filo} Observed shift downwards of the pendulum pole because of a stiff  fishing line. }
\end{figure}

\begin{figure}[!ht]
\centering
\mbox{\epsfig{file=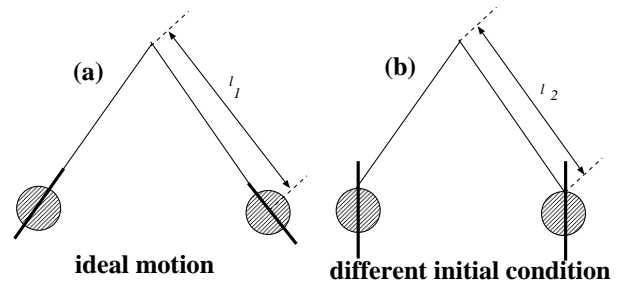,width=8cm}}
\caption{\label{initial-condition} Two different initial conditions of the pendulum motion. A rigid fishing-line can cause the motion as depicted in scheme (a). }
\end{figure}

Results depend on the thread type. In fact, in previous occasions when using a relatively robust fishing line (suitable to be fixed in a deep tiny hole in a block) the average $\bar{g}$ calculated for all groups resulted nearly 1 \% higher than the expected value, considering the length to the center of mass. This result corresponded to a calculated length slightly shorter than the measured one. The problem could not be the period, because as we discussed with the students, after hundreds of records $\bar{T}$ ought to be precise at the per mill. By looking closely at the apparatus during the oscillation while searching for the reasons of the mismatch between the real pendulum length $\ell$ and the length required by calculation according to $\bar{g}$ for all groups, we eventually realized that the stiffer fishing line bent a few millimeters below the pole as shown in Fig.  \ref{piega-filo}
After this observation, the fishing line was changed with a very thin thread, but, unexpectedly, the results remained nearly the same, that is $\bar{g}$ for all groups resulted again higher than expected. We suggested that the ``real'' length $\ell$ was a bit shorter than the measured one because most of the students left the pendulum mass to oscillate in a vertical position instead of a position collinear with the thread as shown in Fig.  \ref{initial-condition}. The object maintains its initial configuration with a zero-rigidity thread. The pendulum length with a vertically positioned mass is slightly shorter than that one in the collinear configuration. These effects, however, are only noticeable if the standard deviations of the measurements are smaller than 0.3-0.4 \%. Obviously, this problem does not show up with the pendulum fixed at the mass-center as in Fig. \ref{pendoli1} (e) . \\

To conclude the first four-hour lab session we introduce the topics of the subsequent session in which the students measure the damping time of the oscillation with two independent methods and compare them. Then we mention other problems such as: 

\begin{itemize}

\item Pendulum motion shows damped oscillations \cite{8}. Data lead to conclude that the damping does not affect the period, i.e. its effect is lower than the measurement errors.

\item Many times the motion is not confined to a plane, it is clearly a composed three-dimensional motion in which both the bob and the pendulum rotate. The pendulum motion comes out much neater with a relatively robust fishing line than with a thread. The (c) and (d) mountings of Fig. \ref{pendoli1} as a bifilar pendulum constrain the motion to a plane \cite{9}. The final {\it g} value is not affected because the causes of error are mainly the measurement of the length and the clock start and stop. In Appendix we report a calculation with a composed motion.

\item Measurements with the thread passing through the center of mass give very precise values.

\item The measurement with a bifilar pendulum gives the same data as a simple pendulum.

\end{itemize}


\section{Measurements of damping time and oscillation decay}

The measurement of the damping constant $\gamma$, that is of the damping time $\tau=1/\gamma$, is done in order to complete the mathematical model of the damped oscillations and, more importantly, this measurement is qualitatively different from the previous ones given that it is based on students' judgment as described below. The time equation is
\begin{equation}
x(t) = x_0 \, e^{-\, \frac{t}{\tau}} \, cos(\omega\, t)
\end{equation}
where the damping time $\tau$ is defined as { \em the time at which the oscillation amplitude is reduced by 1/e}.
\\

\begin{figure}[!ht]
\center
\includegraphics[width=6cm]{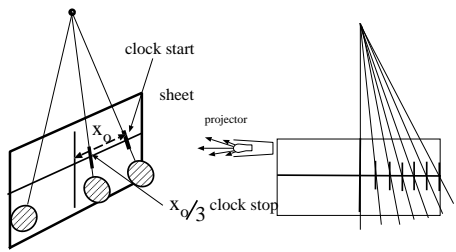}
\caption{\label{damping} Scheme of the damping measurement}
\end{figure}

The measurement set-up for the damping time is shown in Fig. \ref{damping}.
For simplicity we divide the initial amplitude in three parts and consider 1/3 as roughly equivalent to 1/e. 

This measurement has three objectives: (i) to acquaint students with measurements based on their own judgment, therefore giving rise to large errors, (ii) to show how an appropriate technique can make it easy to take an apparently difficult reading, and (iii) to check a physical law by linearising an exponential decay experimental curve.
\\ The set of data are obtained by recording the time interval relative to the different amplitudes, determined by the superposition of the thread shadow and the lines drawn on the screen fixed behind the pendulum as shown in Fig. \ref{damping}.
Typical data are reported in the following table
\vspace{0.2cm}

\begin{tabular}{|c|c|c|}
\hline
group & $\tau$  direct & $\tau$ from fit   \\ \hline
1  &  95  & 93  \\ \hline  
2  & 103   & 137 \\ \hline
3  & 114  &  82  \\ \hline    
4  & 171  & 165  \\ \hline    
5  & 112  & 140 \\ \hline
6  & 85   & 79   \\ \hline    
mean &113 & 116 \\ \hline    
\end{tabular}
\vspace{0.2cm}

The curve $f(t)=exp(-\, t/\tau)$ is transformed into the straight line  
\begin{equation}
 ln\, \left(\frac{x_0}{x(t)}\right) = \frac{1}{\tau}\, t.
\end{equation} 
We discuss with students that it is advisable to organize an experiment with two different methods of measuring a quantity (in order to study the compatibility of the results, and to check for possible systematic errors), and this is especially advisable in our case, in which a value is strongly dependent on the observer's judgment. 
It was also noticed that the five points show a slowly decreasing local slope indicating that the damping time increases as the oscillation amplitude decreases. The slope of this line (the inverse of the damping time) is compared with the value obtained by direct measurement. This experimental observation indicates that the friction (responsible for the damping) increases with velocity. The unexpected success of this session (good compatibility between the data obtained with different experimental paths) rewards students after their hard work.
Data gathered over the years clearly shows higher losses with the relatively rigid fishing lines (confirming the shift of the pole), with bobs oscillating outside the original oscillation plane or with larger oscillation amplitudes.


\section{Revision and further techniques in a new context: the mass-spring experiment}

Given their importance in physics, we have chosen oscillations and waves as the main theme of our laboratory course. After the pendulum, students tackle the mass-spring experiment to revise previous data treatment techniques and to practice the new ones presented in the lectures.

\begin{figure}[!ht]
\centering
\mbox{\epsfig{file=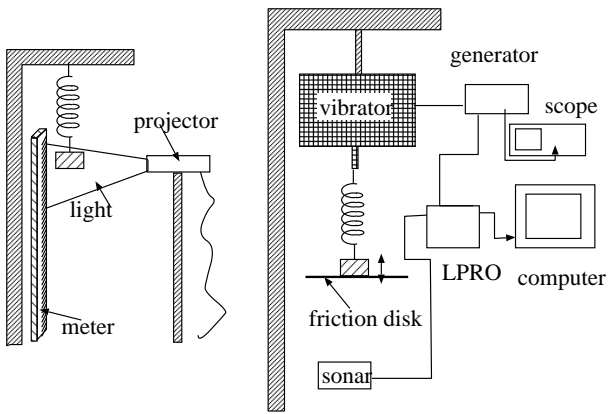,width=8cm}}
\caption{\label{schema-massamolla} Layout scheme for measurements on mass-spring experiment. }
\end{figure}

The layout scheme is shown in Fig.  \ref{schema-massamolla}

We start with the ideal (undamped) motion equation
\begin{equation}
\frac{d^2x}{dt^2} + \omega^2 x = 0 \hspace{2cm} \omega^2=\frac{K}{m}.
\end{equation}
This equation assumes a zero-mass spring, which of course cannot exist in a laboratory. 
Students measure the spring constant {\it K} using both Hooke's law and angular speed $\omega$ measurements according to 
\begin{equation}
\label{hook+omega}
F =  K\, x \hspace{2cm} m = K \, \left(\frac{T}{2\, \pi}\right)^2.
\end{equation}

Comparison of both graphs (\ref{hook+omega}) clearly shows a discrepancy between the two values of {\it K}. To get round of this discrepancy students have to analyze the motion equations in detail. They conclude that theory does not take into consideration the mass of the spring.  \\
Then, as in the previous experiment, the damping time is measured. This result is cross-checked with the full-width-half-maximum (FWHM) of the resonance curve. The equation describing the complete motion is treated theoretically in conjunction with the motion of forced oscillations.  \\
The resonance curve measurements require a substantial upgrade of the experimental apparatus to cause and to detect the oscillations. Students practice on line data acquisition and gather values for the resonance curve. The last part of this rather complex and demanding experiment is the measurement of the phase curve as a function of the frequency of the applied force \cite{10}.


\section{Conclusion}

All measurements and their statistical treatment (gaussian distribution, compatibility of mean values and of standard deviations, curve fitting and confidence level) come out as expected from theory. The technique involved in the experiments is very simple but not trivial. The content in terms of physics is dense because we have to treat the relation between theory and experiment, the weight of approximations, the kind of thread used, and finally the losses dependent on thread rigidity, bob speed and oscillation path.
The content in terms of data processing is equally dense because we have to deal with arithmetic and weighted means, comparison of standard deviations, gaussian distributions of errors, internal compatibility between data and expected results, cross-checking of measurements subject to judgment, test of an exponential decay by means of a straight line fit.
\\
Measuring the gravitational acceleration with the pendulum can be rewarding when the expected result of an important physical quantity is obtained after hours of hard work. Mere solitary data collection can discourage first year students, therefore plenary discussions and subsequent achievement of compatible results in difficult experimental assets can help to reward students, some of which are facing practicals for the first time. 
Following the pendulum experiment with a free and then forced oscillation of a mass-spring experiment gives a practical grounding of SHM theory with a single degree of freedom. By choosing oscillations and waves as a theme for our first year lab course, we provide students with a coherent context in which to practice basic techniques, use of new instruments, and experimental data treatment.

\section{Appendix}

The general motion of a simple pendulum is a composed of the center of mass oscillation and of an oscillatory rotation of the disk around its center of mass.
The system energy has two kinetic components
\[ T_{rotational} = \frac{1}{2} I \, \omega^2 = 
\frac{1}{4} m R^2 \omega^2  \qquad
T_{pendulm} = \frac{1}{2} m \, \ell^2 \dot{\theta}^2.  
\]
Thus the total kinetic energy is
\[ T_{kinetic} = \frac{1}{2} m \, \ell^2 \dot{\theta}^2 + 
\frac{1}{4} m R^2 \omega^2 = \frac{1}{2} m \, \left(\ell^2  \dot{\theta}^2 + 
\frac{1}{2} R^2 \omega^2\right)
\]
In the kinetic energy formula the length of the pendulum with extended mass is the distance from the pole to the center of mass. Since the two angular velocities are equal
$ \dot{\theta} = \omega$
\[T _{kinetic-total} =\frac{1}{2} m \, \left(\ell^2   + 
\frac{1}{2} R^2 \right) \,\theta^2
\]

The potential energy is (using the center of mass)\
\[U = m\,g(\ell - \ell cos\theta) - m\,g \ell \, cos\theta
\]
The lagrangian is
\[ L = \frac{1}{2} m \, \left(\ell^2 +  \frac{1}{2} R^2 \right) \,
\theta^2 + m\,g \ell \, cos\theta 
\]
Given the equation of motion
\[ \frac{d}{dt}\,\frac{\partial L}{\partial \dot{\theta}} -
\frac{\partial L}{\partial \theta} = 0
\]
we deduce
\[ \frac{d}{dt} \left[ m \, \left(\ell^2 +  \frac{1}{2} R^2 \right) \,
\dot{\theta} \right]+ m\,g \ell \, sin\theta =0 
\]

\[ \ddot{\theta} +\frac{g \, sin \theta}{\left (\ell +\frac{R^2}
{2 \ell}\right)} =0 \qquad \textrm{if $\theta$  small}   
\ddot{\theta} +\frac{g \,\theta}{\left (\ell +\frac{R^2}
{2 \ell}\right)} = 0. 
\]
so
\[ \omega = \sqrt{\frac{g}{\ell +\frac{R^2}{2 \ell}}}  \hspace{1cm}
T_{motion} = \frac{2 \pi}{\sqrt{\frac{g}{\ell +\frac{R^2}{2 \ell}}}}
\]
 The rotational motion of the disk reduces the velocity and therefore increases the oscillation time.

\end{document}